\documentclass[aps,noshowkeys,twocolumn,superscriptaddress]{revtex4-1}
\usepackage{epsfig}
\usepackage{graphicx}
\usepackage{color}
\usepackage{bm}
\usepackage{amssymb,amsmath,euscript,esint,mathtools}
\begin{document}
\newcommand{\Br}[1]{(\ref{#1})}
\newcommand{\Eq}[1]{Eq.~(\ref{#1})}
\newcommand{\frc}[2]{\raisebox{1pt}{$#1$}/\raisebox{-1pt}{$#2$}}
\newcommand{\frcc}[2]{\raisebox{0.3pt}{$#1$}/\raisebox{-0.3pt}{$#2$}}
\newcommand{\frccc}[2]{\raisebox{1pt}{$#1$}\big/\raisebox{-1pt}{$#2$}}
\newcommand{\RNumb}[1]{\uppercase\expandafter{\romannumeral #1\relax}}

\title{Progress toward the $\mathcal{P}$, $\mathcal{T}$-odd Faraday effect: Light absorption\\ by atoms briefly interacting with a laser beam}
\author{D. V. Chubukov}
\affiliation{School of Physics and Engineering, ITMO University, Kronverkskiy 49, 197101 St. Petersburg, Russia}
\affiliation{Petersburg Nuclear Physics Institute named by B.P. Konstantinov of National Research Centre ``Kurchatov Institut'', St. Petersburg, Gatchina 188300, Russia}
\author{I. A. Aleksandrov}
\affiliation{Department of Physics, St. Petersburg State University, 7/9 Universitetskaya Naberezhnaya, St. Petersburg 199034, Russia}
\affiliation{Ioffe Institute, Politekhnicheskaya street 26, Saint Petersburg 194021, Russia}
\author{L. V. Skripnikov}
\affiliation{Petersburg Nuclear Physics Institute named by B.P. Konstantinov of National Research Centre ``Kurchatov Institut'', St. Petersburg, Gatchina 188300, Russia}
\affiliation{Department of Physics, St. Petersburg State University, 7/9 Universitetskaya Naberezhnaya, St. Petersburg 199034, Russia}
\author{A.~N.~Petrov}
\affiliation{Petersburg Nuclear Physics Institute named by B.P. Konstantinov of National Research Centre ``Kurchatov Institut'', St. Petersburg, Gatchina 188300, Russia}
\affiliation{Department of Physics, St. Petersburg State University, 7/9 Universitetskaya Naberezhnaya, St. Petersburg 199034, Russia}

\begin{abstract}
We investigate the process of photon absorption by atoms or molecules shortly interacting with a laser beam in the dipole approximation. Assuming that the interaction time $\tau$ is much smaller than the lifetime of the corresponding excited state, we examine the absorption probability as a function of $\tau$. Besides, we incorporate Doppler broadening due to nonzero temperature of the atoms (molecules). It is demonstrated that in the case of a zero detuning and without Doppler broadening, the absorption probability is quadratic in $\tau$. Once Doppler broadening is taken into account or the laser beam is off from the resonant frequency, the absorption probability becomes linear in $\tau$. Our findings are expected to be important for experimental studies in optical cells or cavities where atoms or molecules traverse continuous laser beams. The experimental prospects of searching for the electric dipole moment (EDM) of the electron are discussed in detail.
\end{abstract}
 
\maketitle

\section{Introduction}

A quantum mechanical formalism describing the interaction between light and atoms or molecules was developed as early as in the 1930s (see, e.g., Refs.~\cite{Weiss63,Goep31}). Within quantum electrodynamics (QED), the first approach to the determination of a spectral line profile was proposed by F.~Low~\cite{Low52} (for review, see Ref.~\cite{And08}). An extensive literature has been devoted to laser optics and nonlinear optics since then (see, e.g., Refs.~\cite{Sieg86,Boyd,scully} and references therein). One of the basic phenomena regarding light-matter interactions is the process of photon absorption by atoms or molecules. Here we are interested in evaluation of the excited state population focusing on the setup where the interaction time is much smaller than the lifetime of the excited state of a general two-level system, which is assumed to mimic either an atom or a molecule depending on the experimental scenario.

\begin{figure*}[t]
\center{\includegraphics[width=0.7\linewidth]{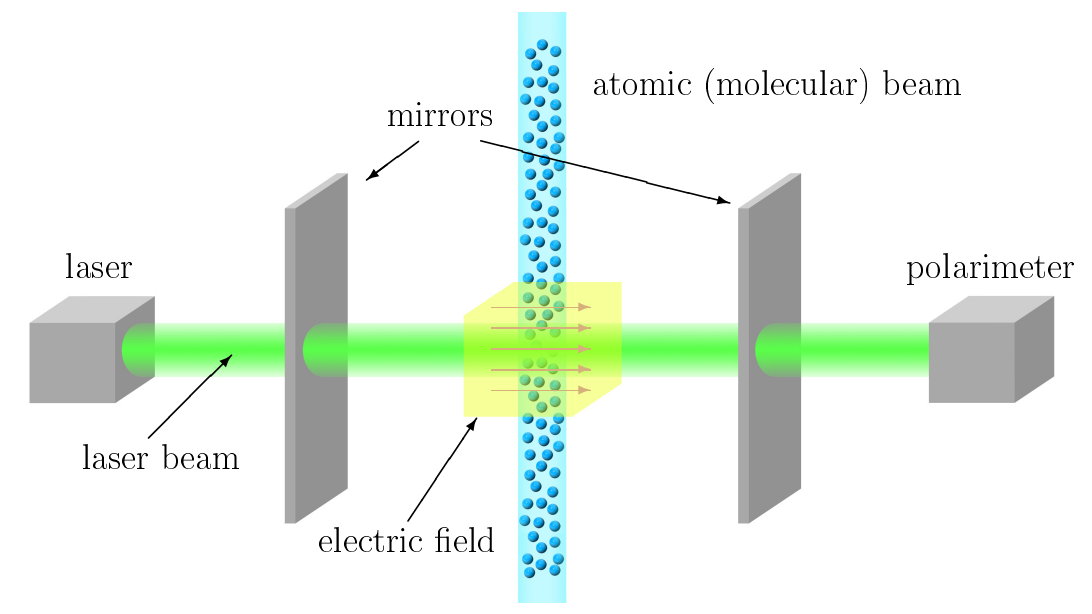}}
\caption{Principle scheme of the experimental setup for measuring the $\mathcal{P}$, $\mathcal{T}$-odd Faraday rotation. A beam of atoms or molecules (blue) traverses a high-finesse optical cavity between two mirrors and interacts with a linearly polarized laser beam (green). The interaction region is located in a background electric field (yellow) directed along the laser beam axis. The laser radiation transmitted through the mirrors is detected by means of a polarimeter.}
\label{fig:scheme}
\end{figure*}

This study is motivated by a very important application of the theory of laser-atom interactions, that is optical experiments where a continuous laser radiation propagates through a gas medium within an optical cell~\cite{Bud02,Khrip91} or high-finesse optical cavity~\cite{Boug14}. The time interval $\tau$ during which the atoms in the gas medium traverse the laser beam (typically 1~mm in diameter), i.e., the so-called ``transit time'', can be smaller than the characteristic lifetime $1/\Gamma$ of the atomic metastable state. This situation arises in the experiment proposed in Refs.~\cite{Chub19,Chub21} for the search for the electric dipole moment (EDM) of the electron. In Refs.~\cite{Chub19,Chub21}, a possibility to reveal such an effect by observing the $\mathcal{P}$, $\mathcal{T}$-odd Faraday rotation ($\mathcal{P}$ stands for the space parity and $\mathcal{T}$ stands for the time reflection invariance) was investigated. The $\mathcal{P}$, $\mathcal{T}$-odd Faraday effect manifests itself as a rotation of the polarization plane of linearly polarized light propagating through a gas medium in an external electric field. The experiment is supposed to be carried out on the ground state of the atomic (or molecular) system with a laser tuned to a resonance with a suitable transition. In this context, it is crucial to avoid large populations of the excited state since it should otherwise substantially reduce the effect hindering the corresponding experimental measurements. This means that one has to require that the absorption probability be sufficiently small. A principle scheme of the proposed experimental setup is depicted in Fig.~\ref{fig:scheme}. An atomic (molecular) beam crosses a high-finesse optical cavity between two mirrors in a transverse direction. Within the cavity, the particles interact with an a linearly polarized laser beam. The overlap of these two beams is located in a background electric field directed along the laser beam axis. The detection of optical rotation (either using simple polarimetry, or phase-sensitive techniques) happens at the output of the cavity. The present constraint (up to recent time) on the electron EDM was established in experiments with ThO molecules by the ACME-collaboration~\cite{ACME18} ($1.1\times 10^{-29}$~$e$cm). Here $e$ stands for the elementary charge. In such an experiment, the electron spin precession in an external electric field was explored. Very recently, a new upper bound for the electron EDM (improvement by a factor $\sim 2.4$) was established in an experiment with trapped molecular ions HfF$^+$ in a rotating electric field~\cite{JILA23}. Still, there is a wide gap between the theoretical predictions of the Standard Model and its extensions and the experimental constraints on the electron EDM (see, e.g., Refs.~\cite{Pos14,Engel2013,Yam21,Posp22}). 
Therefore, it is strongly desirable to develop another type of methods for electron EDM measurements such as that based on the $\mathcal{P}$, $\mathcal{T}$-odd Faraday rotation. In the case of the ACME-type experiment, the statistics is determined by the number of molecules, whereas within the $\mathcal{P}$, $\mathcal{T}$-odd Faraday experiment, the statistics can be governed by the number of detected photons. In the latter case, one has to conduct the experiment using atoms or molecules in a ground state with a zero total angular momentum. Note that the excited state can be considered optically inactive for the laser, provided the decoherence effects are negligible. In this case, there will be no noise induced by atoms or molecules. In Refs.~\cite{Chub19,Chub21} the $\mathcal{P}$, $\mathcal{T}$-odd Faraday experiment on ThO and PbF was proposed. The ground state of the ThO molecule possesses a zero total angular momentum. The PbF molecule with account for the nuclear spin can also be formally considered to have a zero total angular momentum~\cite{Maw11,Pet13}. It is important that one can make the number of detected photons (that had interacted with molecules and scattered forward) much larger than molecules. Then, our aim is to have high laser intensity keeping the absorption probability much less than unity. A rough estimate of the absorption probability was used in Ref.~\cite{Chub21}. In the present study, we thoroughly examine this quantity in the dipole approximation taking into account the laser detuning from the atomic resonance and Doppler broadening.

We employ the relativistic units $\hbar=c=1$ ($\hbar$ is the Planck constant, $c$ is the speed of light). The charge units correspond to $\alpha = e^2/(4 \pi)$ ($e<0$ is the electron charge, $\alpha$ is the fine-structure constant). Throughout the paper, we will refer to the two-level system as the ``atom'' although it may well also be a molecule. Also we do not distinguish between the terms ``transit time'' and ``interaction time''.

\begin{figure}[b]
\center{\includegraphics[width=0.9\linewidth]{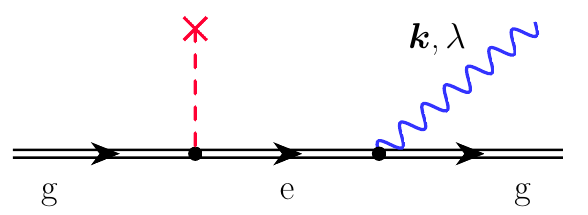}}
\caption{Leading-order Feynman diagram describing absorption and emission of a photon by an atom interacting with a classical external field.}
\label{fig:diagram}
\end{figure}

\section{Calculation of the absorption probability} \label{sec:calc}

Within the conventional formalism of QED, the initial and final (asymptotic) states are eigenstates of the Hamiltonian incorporating the Coulomb interaction. In the case of an atom, these can only correspond to the ground state $\psi_\text{g}$ since the excited state has a finite lifetime, i.e., nonzero natural width $\Gamma$. Accordingly, the absorption process must be followed by photon emission. Nevertheless, in the stationary case (infinite duration of the interaction time between the atom and external field) these two stages can be considered separately, i.e., the corresponding Feynman diagram with two vertices can be factorized. However, it is no longer applicable if the atom interacts with a time-dependent background radiation, e.g., laser field. In what follows, we will assume that the interaction time $\tau$ is small so that $\Gamma \tau \ll 1$. The laser field is treated as a classical external background, so within perturbation theory, the absorption-emission process is described by the Feynman diagram depicted in Fig.~\ref{fig:diagram}~\cite{LL4,Schweber}. The double line represents the exact electron wavefunction (or propagator) incorporating the interaction with the nucleus and other electrons, whereas the laser field is treated perturbatively. The latter approximation is well justified as we are interested in the (near-)resonance case, i.e., the laser frequency $\omega_\text{L}$ is close to the energy difference $\omega_0=E_1-E_0$ ($E_0$ and $E_1$ are the energies of the ground and excited states, respectively). The diagram yields the main contribution when the intermediate state coincides with the excited state $\psi_\text{e}$. Note also that the diagram with the different order of the absorption and emission events is strongly suppressed in the resonance case, so it will be disregarded. The condition $\Gamma\tau \ll 1$ prevents reabsorption, so it is the diagram in Fig.~\ref{fig:diagram} that contains all the necessary information and there is no need in taking into account the higher-order diagrams.

Integrating the mod-square of the diagram over the photon momentum and summing over the polarizations, one obtains the absorption probability $W$. From the practical viewpoint, it is clear that photon emission takes place after the interaction with the laser field as $\tau \ll 1/\Gamma$, so $W$ can also be interpreted as the population of the excited state after the interaction, provided $W \ll 1$. The latter condition is actually one of particular interest since our goal is to keep the excited-state population small. Finally, we note that as long as $W \ll 1$, the stimulated emission process is suppressed.

Let us evaluate the amplitude $S_{\boldsymbol{k},\lambda}$ of the process depicted in Fig.~\ref{fig:diagram}. The general expression reads
\begin{equation}
 \label{amp_0}
S_{\boldsymbol{k},\lambda}=e^2\int d^4 x d^4y \overline{\psi}_\text{g}(y) \hat{\varepsilon}^{\ast}_{\boldsymbol{k},\lambda}\Phi^{\ast}_{\boldsymbol{k},\lambda}(y) S(y,x) \hat{\mathcal{A}}(x) \psi_\text{g}(x),
\end{equation}
where $\overline{\psi} \equiv \psi^\dagger \gamma^0$, $\hat{V} \equiv \gamma^{\mu}V_{\mu}$, $\varepsilon^{\mu}_{\boldsymbol{k},\lambda}$ is the photon polarization four-vector (the photon momentum and polarization are denoted by $\boldsymbol{k}$ and $\lambda$, respectively), $\Phi_{\boldsymbol{k},\lambda}$ is the space-time-dependent part of the photon wavefunction, $S(y, x)$ is the bound electron propagator, and $\mathcal{A}^\mu$ is a four-vector describing the laser field and involving the interaction time $\tau$. Let us introduce the following temporal and spatial components: $x = \{ t_x, \boldsymbol{r}_x \}$ and $y = \{ t_y, \boldsymbol{r}_y \}$. The ground state wavefunction has the form
\begin{equation}
 \label{ground}
\psi_\text{g} (x) = \psi_\text{g} (\boldsymbol{r}_x) {\rm e}^{-iE_0 t_x}.
\end{equation}
The function $\Phi_{\boldsymbol{k},\lambda}$ reads
\begin{equation}
    \label{ph_wf}
 \Phi_{\boldsymbol{k},\lambda} (x)= \frac{1}{\sqrt{2k_0}} \, {\rm e}^{-i(k_0t_x-\boldsymbol{k}\boldsymbol{r}_x)},
\end{equation}
where $k_0 = |\boldsymbol{k}|$. The electron propagator can be represented in the following form:
\begin{equation}
    \label{e_propagator}
 S(y,x)= \int \frac{d\tilde{\omega}}{2\pi}{\rm e}^{-i\tilde{\omega}(t_y-t_x)} \sum_n \frac{\psi_n(\boldsymbol{r}_y)\overline{\psi}_n(\boldsymbol{r}_x)}{\tilde{\omega}-E_n(1-i0)},
\end{equation}
where the sum is taken over the whole spectrum of the system, $E_n$ is the energy of the state $\psi_n$. Since we consider a two-level system and resonance effects, only one term of the sum corresponding to the resonance excited state $\psi_n=\psi_{\text{e}}$ survives. The explicit form of the wavefunction is
\begin{equation}
    \label{excited}
\psi_{\text{e}} (x)  = \psi_{\text{e}} (\boldsymbol{r}_x) {\rm e}^{-iE_1 t_x}.
\end{equation}
The classical laser field is described by the following vector potential:
\begin{equation}
\boldsymbol{\mathcal{A}}(x)= \frac{\mathcal{E}_{\rm L}}{\omega_{{\rm L}}} \boldsymbol{e}_{\rm L}{\rm e}^{-i(\omega_{{\rm L}}t_x-\boldsymbol{k}_{\rm L}\boldsymbol{r}_x)}{\rm e}^{-4t_x^2/\tau^2},
\label{las_wf}
\end{equation}
where $\mathcal{E}_{\rm L}$ is the field strength, and $\omega_{\rm L}=|\boldsymbol{k}_{\rm L}|$. The scalar potential vanishes, $\mathcal{A}_0 = 0$. The last exponential function in Eq.~\eqref{las_wf} introduces a finite interaction time. The transverse profile of the laser field along, say, the $x$ direction is given by $\mathrm{exp} (-x^2/w^2)$, where $w$ governs the laser beam radius. In this slowly varying envelope, we replace $x$ with the classical expression $v_x t_x$, where $v_x$ is the average speed of the atoms. Introducing then $\tau = 2w/v_x$, one arrives at Eq.~\eqref{las_wf}. Note that the rapidly oscillating carrier exponent is treated exactly, i.e., we incorporate its spatial dependence.

Integrating over the temporal variables $t_x$ and $t_y$ in \Eq{amp_0}, one obtains
\begin{widetext}
\begin{equation}
    \label{amp_1}
S_{\boldsymbol{k},\lambda}= \frac{\sqrt{2} \pi^{3/2}\alpha\tau\mathcal{E}_{\rm L} }{\omega_{\rm L} \sqrt{k_0}} \langle {\rm g} | \hat{\varepsilon}^{\ast}_{\boldsymbol{k},\lambda} {\rm e}^{-i\boldsymbol{k}\boldsymbol{r}} | {\rm e}\rangle \langle {\rm e} | \hat{\boldsymbol{e}}_{\rm L} {\rm e}^{i\boldsymbol{k}_{\rm L}\boldsymbol{r}} |  {\rm g}\rangle \frac{1}{k_0-\omega_0+ i\Gamma/2} \, {\rm e}^{-(k_0-\omega_{\rm L})^2 \tau^2 /16},
\end{equation}
where $\hat{\boldsymbol{e}}_{\rm L} \equiv - \boldsymbol{\gamma} \boldsymbol{e}_{\rm L}$ and the spatial matrix elements are defined via
\begin{equation}
\langle n | f(\boldsymbol{r}) | m \rangle \equiv \int d\boldsymbol{r} \overline{\psi}_n(\boldsymbol{r}) f(\boldsymbol{r}) \psi_m(\boldsymbol{r}).
\end{equation}
In Eq.~\eqref{amp_1} we performed the conventional substitution $E_1 \to E_1 - i\Gamma/2$ in order to regularize the denominator. The natural width $\Gamma$ appears due to the radiative corrections (mass operator) to the electron wavefunction and determines the lifetime of the excited state (the standard derivation can be found, e.g., in Refs.~\cite{LL4, Andreev08}).

The probability density of the process (probability $W_{\lambda}$ per unit phase volume $d^3\boldsymbol{k}$) is given by
\begin{equation}
    \label{dens_prob}
\frac{dW_{\lambda}}{d^3\boldsymbol{k}}= \frac{|S_{\boldsymbol{k},\lambda}|^2}{(2\pi)^3}.
\end{equation}
Taking the mod-square of $S_{\boldsymbol{k},\lambda}$ in \Eq{amp_1}, we obtain 
\begin{equation}
    \label{prob_1}
W = \sum_\lambda W_\lambda = \frac{\alpha^2\tau^2\mathcal{E}_{\rm L}^2 }{4\omega_{\rm L}^2} \Big|\langle {\rm e} | \hat{\boldsymbol{e}}_{\rm L} {\rm e}^{i\boldsymbol{k}_{\rm L}\boldsymbol{r}} |  {\rm g}\rangle\Big|^2 \sum_{\lambda} \int \frac{d^3\boldsymbol{k}}{k_0}  \Big|\langle {\rm g} | \hat{\varepsilon}^{\ast}_{\boldsymbol{k},\lambda} {\rm e}^{-i\boldsymbol{k}\boldsymbol{r}} | {\rm e}\rangle\Big|^2 \frac{{\rm e}^{-(k_0-\omega_{\rm L})^2 \tau^2 /8}}{(k_0-\omega_0)^2+ \Gamma^2/4}.
\end{equation}
\end{widetext}
Let us now perform the summation over the photon polarization $\lambda$. The polarization four-vector has the form $\varepsilon_{\boldsymbol{k},\lambda} = (0, \boldsymbol{e}_{\boldsymbol{k},\lambda})$. To calculate the matrix element in Eq.~\eqref{prob_1}, we approximately replace ${\rm e}^{-i\boldsymbol{k}\boldsymbol{r}}$ with $1$ and identify the matrix element of $\boldsymbol{\gamma}$ with the electron velocity according to the nonrelativistic approximation. After that, we employ the following identity, which is valid for any vector $\boldsymbol{a}$~\cite{LL4}: $\sum_{\lambda} |\boldsymbol{e}_{\boldsymbol{k}, \lambda} \boldsymbol{a}|^2=|\boldsymbol{n} \times \boldsymbol{a}|^2 $, where $\boldsymbol{n} \equiv \boldsymbol{k}/k_0$. The matrix element of the velocity is proportional to the matrix element of the position vector, so one finally arrives at
\begin{equation}
    \label{dipole_emit}
\sum_{\lambda}   \Big|\langle {\rm g} | \hat{\varepsilon}^{\ast}_{\boldsymbol{k},\lambda} {\rm e}^{-i\boldsymbol{k}\boldsymbol{r})} | {\rm e}\rangle\Big|^2\approx \omega_0^2 \big|\boldsymbol{n}\times \boldsymbol{r}_{{\rm ge}} \big|^2.
\end{equation}
Here $\boldsymbol{r}_{{\rm ge}}=\langle {\rm g} | \boldsymbol{r} | {\rm e}\rangle$.
Note that the matrix element of the electric dipole moment is $\langle {\rm g} | \boldsymbol{d} | {\rm e}\rangle=\boldsymbol{d}_{\rm ge} = e \boldsymbol{r}_{\rm ge}$. Analogous calculations yield
\begin{equation}
    \label{dipole_abs}
\Big|\langle {\rm e} | \hat{\boldsymbol{e}}_{\rm L}{\rm e}^{i\boldsymbol{k}_{\rm L}\boldsymbol{r})} |  {\rm g}\rangle\Big|^2 \approx \omega_0^2 \big| \boldsymbol{e}_{\rm L}\boldsymbol{r}_{{\rm eg}} \big|^2,
\end{equation}
where $\boldsymbol{r}_{{\rm eg}}=\langle {\rm e} | \boldsymbol{r} | {\rm g}\rangle$. Next we will integrate over the emitted photon momentum $\boldsymbol{k}$, which can be parametrized by the energy $k_0$ and two spherical angles. If the $k_z$ axis is directed along the vector $\boldsymbol{r}_{{\rm ge}}$, then $|\boldsymbol{n}\times \boldsymbol{r}_{{\rm ge}}|^2= |\boldsymbol{r}_{{\rm ge}}|^2 \sin^2\theta$ and the integration over the angles yields
\begin{equation}
    \label{prob_2}
W= \frac{16\pi^2}{3} \, \alpha^2\tau^2 \omega_{\rm L}^2I \big| \boldsymbol{e}_{\rm L}\boldsymbol{r}_{{\rm eg}} \big|^2 \big|\boldsymbol{r}_{{\rm ge}}\big|^2 \int\limits_0^\infty \frac{k_0 {\rm e}^{-(k_0-\omega_{\rm L})^2\tau^2/8}dk_0}{(k_0-\omega_0)^2+ \Gamma^2/4},
\end{equation}
where $I=\mathcal{E}_{\rm L}^2/(8\pi)$ is the laser intensity. In the overall prefactor in \Eq{prob_2}, it is assumed that $\omega_{\rm L}\approx \omega_0$. An explicit evaluation of the integral in \Eq{prob_2} is carried out in the Appendix. The result reads
\begin{equation}
    \label{prob3}
W= \frac{32\pi^{5/2}}{3} \frac{\alpha^2\tau^2 \omega_{\rm L}^3I }{\Gamma}  \big| \boldsymbol{e}_{\rm L}\boldsymbol{r}_{{\rm eg}} \big|^2 \big|\boldsymbol{r}_{{\rm ge}}\big|^2 f\left(\frac{\Delta\omega\tau}{2\sqrt{2}},\frac{\Gamma\tau}{4\sqrt{2}}\right).
\end{equation}
Here $\Delta \omega=\omega_{\rm L}-\omega_0$ is a detuning from the atomic resonance, and the real-valued function $f(x,y)$ is defined via
\begin{equation}
\label{fg}
f(x,y)= {\rm Re} \, \sqrt{\pi} {\rm e}^{-z^2}[1-\text{erf}(-iz)],
\end{equation}
where $z=x+iy$. It is natural to expect that the excited state population of the atom should not depend on the emission matrix element. Actually, the dipole matrix element can be expressed via the natural width as follows (see, e.g., Ref.~\cite{LL4}):
\begin{equation}
\label{rge}
\big|\boldsymbol{r}_{{\rm ge}}\big|^2=\frac{3\Gamma}{4\alpha\omega_0^3}\approx \frac{3\Gamma}{4\alpha\omega_{\rm L}^3}.
\end{equation}
The expression~\Br{rge} (dipole approximation) is accurate since the light wavelength is much greater than the Bohr radius. Then, we rewrite \Eq{prob3} in the form
\begin{equation}
    \label{prob4}
W= 8\pi^{5/2} \alpha\tau^2I  \big| \boldsymbol{e}_{\rm L}\boldsymbol{r}_{{\rm eg}} \big|^2  f\left(\frac{\Delta\omega\tau}{2\sqrt{2}},\frac{\Gamma\tau}{4\sqrt{2}}\right).
\end{equation}
In the case of a zero detuning ($\Delta \omega=0$) and $\Gamma \tau \ll 1$, we have $f(0,\Gamma \tau/4\sqrt{2}) \approx \sqrt{\pi}$, so one obtains a quadratic dependence of the absorption probability $W_0$ on the transit time~$\tau$:
\begin{equation}
    \label{zero_det}
W_0\approx 8\pi^3\alpha\tau^2I  \big| \boldsymbol{e}_{\rm L}\boldsymbol{r}_{{\rm eg}} \big|^2.
\end{equation}
This result can be interpreted in the following way. Due to the short transit time ($\Gamma\tau \ll 1$), the absorption lineshape gains a frequency profile with effective width $\Gamma_\tau\sim 1/\tau$. Then, at the exact resonance, the effect is determined by the largest of the widths, i.e., the absorption probability is proportional to $\tau/\Gamma_\tau\sim \tau^2$ (compare with the standard expression for the stationary probability per unit time in the resonance $= 1/\Gamma$~\cite{Sieg86}).
Accordingly, the absorption probability rapidly decreases for short-time interactions (although for large $\tau$, the probability may formally exceed unity, so one has to take into account the higher-order contributions with respect to the interaction with the classical background, here we neglect the higher-order terms since in our analysis the probability is always small). Within the setup considered in Refs.~\cite{Chub19,Chub21}, the transit time inside the cavity was assumed to be $10^{-5}$~s, while the natural width of the molecular state excited by the laser radiation amounted to $\Gamma \sim 10^3~\text{s}^{-1}$. This may represent a favorable scenario for reaching a higher accuracy of measurements. Nevertheless, for experiments proposed in Refs.~\cite{Chub19,Chub21}, it is also of great importance to have large detunings $\Delta \omega$, which will be discussed next.

The asymptotic behavior of $f(x,y)$ for $x\gg 1$ and $y \ll 1$ is given by
\begin{eqnarray}
f(x,y) &=& \frac{y}{x^2}+\mathcal{O}(x^{-4}) \label{asym_f}.
\end{eqnarray}
Note that this expression differs from \Eq{fg} by no more than 10 percent already for $x \gtrsim 4$~\cite{Chub19,Chub21}, which is sufficient for our aims.
Substituting now Eq.~\eqref{asym_f} into
\Eq{prob4}, we obtain the following expression for the absorption probability in the case $\Delta \omega \tau \gg 1$ in the limit of small $\tau$ ($\Gamma \tau \ll 1$):
\begin{equation}
    \label{large_det}
W_{\Delta}\approx \frac{2^{7/2} \pi^{5/2} \alpha\Gamma\tau I}{(\Delta\omega)^2} \, \big| \boldsymbol{e}_{\rm L}\boldsymbol{r}_{{\rm eg}} \big|^2.
\end{equation}
Thus, the leading contribution to the absorption probability within the small-$\tau$ regime for large detunings is linear in the transit-time parameter. We evaluated the absorption probability in the case of an individual atom. Next, we will consider an atomic ensemble at finite temperature.

\section{Doppler broadening} \label{sec:doppler}

In our calculations, we should address a further broadening mechanism which affects the lineshape --- Doppler broadening. We assume that the atoms have the same longitudinal velocity, so the transit time $\tau$ is well defined, while the transverse velocity fluctuates (the atoms can slowly move along the laser beam axis). We introduce Doppler broadening via a convolution of \Eq{prob_2} with the Maxwell distribution for the atoms at given temperature $T$. This convolution represents the averaging over the chaotic motion of the atoms.
Let us denote the probability in Eq.~\eqref{prob_2} by $W(\tau, \omega_\text{L})$. We perform the averaging via
\begin{equation}
    \label{doppler}
    \overline{W}(\tau, \omega_{\text{L}}) =\int \limits^{\infty}_{-\infty}dv \, W(\tau,\omega_{\rm L}-v\omega_{\rm L})P(v),
\end{equation}
where $v$ is the projection of atomic velocity on the laser beam direction, $P(v)$ is a one-dimensional Maxwell distribution function,
\begin{equation}
    \label{14}
    P(v)=\frac{1}{\sqrt{\pi} v_0} \, {\rm e}^{-v^2/v_0^2}, \qquad v_0=\sqrt{\frac{2kT}{M}}.
\end{equation}
Here $k$ is the Boltzmann constant and $M$ is the mass of the atom. The Doppler width is defined as $\Gamma_{\text{D}}=\omega_0 v_0\approx \omega_{\text{L}} v_0$ ($\Delta\omega \ll \omega_0$). Introducing a dimensionless variable $x=v/v_0$, we recast \Eq{doppler} into
\begin{widetext}
\begin{equation}
    \label{prob_dop}
    \overline{W}(\tau, \omega_{\text{L}}) =  \frac{4 \pi^{3/2}\alpha\Gamma\tau^2 I \big| \boldsymbol{e}_{\rm L}\boldsymbol{r}_{{\rm eg}} \big|^2}{\omega_{\rm L}}\int \limits^{\infty}_{-\infty}dx \, {\rm e}^{-x^2} \int\limits_0^\infty \frac{k_0 {\rm e}^{-(k_0-\omega_{\rm L}+\Gamma_{\rm D}x)^2\tau^2/8}}{(k_0-\omega_0)^2+ \Gamma^2/4} \, dk_0.
\end{equation}
In what follows, we will assume $\Gamma_{\text{D}} \tau \gg 1$, which represents a completely realistic condition~\cite{Chub19,Chub21}. To carry out the integration in \Eq{prob_dop}, let us first evaluate the integral over $x$.  Neglecting exponentially suppressed terms as $\Gamma_{\text{D}} \tau \gg 1$, one obtains
\begin{equation}
    \label{int_x}
   \int \limits^{\infty}_{-\infty}dx \, {\rm e}^{-x^2} {\rm e}^{-(k_0-\omega_{\rm L}+\Gamma_{\rm D}x)^2\tau^2/8}\approx\frac{\sqrt{8\pi}}{\Gamma_{\rm D}\tau} \, {\rm e}^{-(k_0-\omega_{\rm L})^2/\Gamma_{\rm D}^2} .
    \end{equation}
Then, the leading-order contribution in $\tau $ in \Eq{prob_dop} takes the form
\begin{equation}
    \label{prob_dop2}
    \overline{W}(\tau, \omega_{\text{L}}) \approx  \frac{2^{7/2} \pi^{2}\alpha\Gamma\tau I \big| \boldsymbol{e}_{\rm L}\boldsymbol{r}_{{\rm eg}} \big|^2}{\omega_{\rm L}\Gamma_{\rm D}} \int\limits_0^\infty \frac{k_0 {\rm e}^{-(k_0-\omega_{\rm L})^2/\Gamma_{\rm D}^2}}{(k_0-\omega_0)^2+ \Gamma^2/4} \, dk_0.
\end{equation}
\end{widetext}
The integral in \Eq{prob_dop2} has the same form as in \Eq{prob_2}. It yields
\begin{equation}
    \label{prob_dop3}
    \overline{W}(\tau, \omega_{\text{L}}) =  \frac{2^{9/2} \pi^{5/2}\alpha\tau I \big| \boldsymbol{e}_{\rm L}\boldsymbol{r}_{{\rm eg}} \big|^2}{\Gamma_{\rm D}}f(u,q),
    \end{equation}
where $u=\Delta\omega/\Gamma_{\rm D}$ and $q=\Gamma/(2\Gamma_{\rm D})$.  One observes that in the leading order the $\tau$ dependence and the form of the expression coincide with a rough estimate given in Ref.~\cite{Chub21}. Note also that compared to the non-Doppler case, now for a zero detuning one has a linear dependence on the transit time since the largest of the widths is the Doppler one ($\overline{W}$ is now proportional to $\tau/\Gamma_\text{D}$ instead of $\tau/\Gamma_\tau$).

In what follows, we will consider the regime of large detunings ($u=4$--$5$). Utilizing the asymptotic behavior of $f(u,q)$ [see \Eq{asym_f}], one obtains from \Eq{prob_dop3}
\begin{equation}
    \label{doppler_large_det}
    \overline{W}_{\Delta}(\tau, \omega_{\text{L}}) \approx   \frac{2^{7/2} \pi^{5/2}\alpha\Gamma\tau I \big| \boldsymbol{e}_{\rm L}\boldsymbol{r}_{{\rm eg}} \big|^2}{u^2\Gamma^2_{\rm D}}.
\end{equation}
Averaging over the atomic ensemble, we replace $\big| \boldsymbol{e}_{\rm L}\boldsymbol{r}_{{\rm eg}} \big|^2$ with $\big| \boldsymbol{r}_{{\rm eg}} \big|^2/2= \big| \boldsymbol{r}_{{\rm ge}} \big|^2/2$. Then, using \Eq{rge}, we rewrite \Eq{doppler_large_det} in the following form:
\begin{equation}
    \label{doppler_large_det2}
    \overline{W}_{\Delta}(\tau, \omega_{\text{L}}) \approx  \frac{3\sqrt{2} \pi^{5/2}\Gamma^2\tau I}{u^2\Gamma^2_{\rm D}\omega_{\text{L}}^3}.
\end{equation}
This result suggests that there is no significant reduction of the absorption probability due to short laser-atom interactions in the case of large detunings. However, the advantage of the scenario with the transit time $\tau\ll 1/\Gamma$ over the stationary one can be demonstrated in the following way. In the stationary case, when one increases the laser intensity, at a certain point the refractive index reduces to a half compared to the system without the laser field. This corresponds to the so-called saturation intensity $I_{\rm sat}$, for which the rates of absorption, stimulated emission and spontaneous decay become equal. According to Ref.~\cite{Sieg86}, for large detunings the saturation intensity reads
\begin{equation}
    \label{sat_int}
   I_{\rm sat}= \frac{2\omega_{\rm L}^3 (\Delta\omega)^2}{\pi\Gamma}.
\end{equation}
Substituting the expression for the saturation intensity \eqref{sat_int} in \Eq{doppler_large_det2}, one obtains
\begin{equation}
    \label{prob_sat}
   \left. \overline{W}_{\Delta}(\tau, \omega_\text{L})\right|_{I=I_\text{sat}} = 6\sqrt{2} \pi^{3/2} \Gamma\tau \approx 50 \Gamma\tau .
\end{equation}
Thus, we justified the rough parametric estimate $\left. \overline{W}_{\Delta}(\tau, \omega_\text{L})\right|_{I=I_\text{sat}} \approx \Gamma\tau$ that was used in Ref.~\cite{Chub21}. From \Eq{prob_sat} it follows that $\left. \overline{W}_{\Delta}(\tau, \omega_\text{L})\right|_{I=I_\text{sat}} \ll 1$ for sufficiently small $\Gamma \tau$, while in the stationary case this quantity amounts to $1/3$.

\section{$\mathcal{P}$, $\mathcal{T}$-odd Faraday effect}

Here we apply the results obtained above to the proposed optical cavity experiment concerning the observation of the $\mathcal{P}$, $\mathcal{T}$-odd Faraday effect. We consider an experiment with heavy heteronuclear molecules rather than with atoms since it is known that the $\mathcal{P}$, $\mathcal{T}$-odd effects in molecules are additionally enhanced due to the presence of the $\Omega$-doubling effect~\cite{Lab78,Sushkov78,Gor79}. This manifests itself in a splitting of every electron energy level into two very close sublevels with opposite spatial parities.

Let us compare the experimental setup involving a molecular beam and that based on using a molecular vapor (medium). First, in a medium, the collisional width $\Gamma_{\text{col}}$ is usually larger than the natural one, which diminishes the rotation angle, while for the beam case $\Gamma_{\text{col}}$ can be neglected. Second, if a laser beam intersects a molecular one at right angle, then the transverse Doppler width should be taken into account. However, it is usually smaller than the ordinary one, which is also favorable for the considered effect. Third, we note that a molecular vapor usually interacts chemically, reacts with walls, settles on walls, etc. Moreover, there are technical difficulties in creating a large external electric field over a long distance. For instance, a PbF molecule (a good candidate for the observation of the $\mathcal{P}$, $\mathcal{T}$-odd effects) is polarized in an electric field $\mathcal{E} \sim 10^4$~V/cm. Such a large field can be arranged only within the distance of several centimeters. On the other hand, the typical diameter of the beam (several centimeters) is smaller than the typical cavity length (about a meter), which obviously diminishes the resulting rotation angle. We believe that this drawback of the molecular-beam setup is minor compared to the advantages described above. Finally, as will be shown below, the transit time effects, which represent the main subject of the present paper, also enhances the effect from the statistical viewpoint. In what follows, we will focus on discussing the experiment concerning optical rotation with a molecular beam.

Let us discuss the main notations and expressions for estimating the experimental signal. The rotation signal $R(u)$ reads
\begin{equation}
    \label{rot_signal}
  R(u) = \psi(u) N_{\text{ev}},
\end{equation}
where $u = \Delta \omega /\Gamma_\text{D}$ is a dimensionless detuning introduced in Section~\ref{sec:doppler}, $\psi(u)$ is the rotation angle and $N_{\text{ev}}$ is the number of statistical events which, in our case, is the number of photons that interacted with molecules (scattered forward) and then were detected. The transmission function $T(u,q)$ related to the intracavity losses obeys the Beer-Lambert law:
\begin{equation}
\label{Beer_law}
T(u,q)=\frac{I(u,q)}{I_0}=\frac{N_{\text{ev}}}{N_0}=\mathrm{e}^{-l /L(u,q)},
\end{equation}
where $I(u,q)$ is the final detected intensity, $I_0$ is the initial intensity, $N_0$ is the initial number of photons emitted by the laser, and $L(u,q)$ is the absorption length. The dimensionless parameter $q = \Gamma/(2\Gamma_\text{D})$ was defined in the previous Section, and
\begin{equation}
\label{abs_length}
L(u,q)= [\rho\sigma(u,q)]^{-1}=[\rho\sigma_0 f(u,q)]^{-1}.
\end{equation}
Here $\rho$ is the number density of molecules and $\sigma_0$ is the standard resonance cross-section for the photon absorption by a molecule; the function $f(u,q)$ is defined in~\Eq{fg}. Being expressed via the absorption length at the given detuning, the rotation signal reads~\cite{Chub21}
 \begin{equation}
 \label{rot_sign_2}
R (u)  =  \frac{h(u,q)}{f(u,q)} \frac{l}{L(u,q)}  \frac{d_e  \mathcal{E}_{\text{eff}}}{2\Gamma_{\text{D}}}N_{\text{ev}}.
\end{equation}
The real-valued function $h(u,q)= \frac{\partial}{\partial u} g(u,q)$, where the function $g(u,q)$ is defined in~\Eq{faddeeva_int} (see also Ref.~\cite{Khrip91}). The quantity $l$ is the optical pathlength, i.e. the total pathlength where laser light interacts with a molecular beam before detection. For instance, for the cavity with the reported transmission $\delta\sim 10^{-5}$~\cite{Boug14} crossed by a molecular beam (typically 1 cm in diameter), it is $l\sim 10^5$~cm. Another impressive achievement is that the cavity with $\delta\sim 10^{-8}$--$10^{-7}$~\cite{Baev99} yields $l\sim 10^7-10^8$~cm for laser light propagating through a molecular beam. Besides, in~\Eq{rot_sign_2}, $d_e$ is the value of the electron EDM, whose constraint is established in such experiments, i.e., the corresponding experiments provide an access, if indirect, to estimating the electron EDM. Finally, $\mathcal{E}_{\text{eff}}$ is the internal molecular effective electric field acting on the electron EDM, which cannot be measured experimentally and should be calculated (see, e.g., Refs.~\cite{Tit06,Skrip17}).

In the previous Sections, we derived general expressions allowing one to address both resonance case and large detunings. Next, we will discuss and compare these two experimental options and demonstrate that using a large-detuning setup is more advantageous. At the resonant frequency $h(0,q)\sim f(0,q)\sim 1$. Then, \Eq{rot_sign_2} can be rewritten as
\begin{equation}
\label{rot_sign_0}
R (u=0)  \approx \frac{l}{L(0,q)}  \frac{d_e  \mathcal{E}_{\text{eff}}}{2\Gamma_{\text{D}}}N_{\text{ev}}.
\end{equation}
The large-detuning asymptotics reads $|h(u,q)|\approx 1/u^2+\mathcal{O}(u^{-4})$, while that of $f(u,q)$ is presented in \Eq{asym_f}. As was stated in Section~\ref{sec:calc}, these asymptotics are valid within 10 percent already for $u \gtrsim 4$~\cite{Chub19,Chub21}. Then, large-detuning rotation signal reads
\begin{equation}
\label{rot_sign_large}
R (u\gtrsim 4)  \approx \frac{l}{L(u,q)}  \frac{d_e  \mathcal{E}_{\text{eff}}}{\Gamma}N_{\text{ev}}.
\end{equation}
As was mentioned in the Introduction and showed in Ref.~\cite{Chub21}, for the observation of the $\mathcal{P}$, $\mathcal{T}$-odd Faraday effect in optical cavities, the fundamental photon shot-noise determines the statistical error of the experiment (not number of molecules). Note that in this case, molecules with zero total angular momentum should be used. It means that the more injected and, accordingly, detected photons per second we have, the better statistical sensitivity of the experiment is. Then, one can define the noise $F$ as $F=\sqrt{N_{\text{ev}}}$, where $N_{\text{ev}}$ is the number of detected photons. For the shot-noise limited measurement, one should keep the signal-to noise ratio greater than unity:
\begin{equation}
\label{R-to-F}
\frac{R(u)}{F}=\psi(u)\sqrt{N_{\text{ev}}}>1.
\end{equation}
The vital point here is that the number of photons $N_{\text{ev}}$ is limited by the maximum possible intracavity intensity which in turn is determined by the intensity that saturates the transition (in other words, ``bleaches'' the molecules).

Let us now discuss a standard optical optimization for both resonance and off-resonance setups. This procedure requires that the signal-to-noise ratio~\Br{R-to-F} be maximized over the optical pathlength $l$. Here we employ Eq.~\Br{Beer_law} for $N_\text{ev}$ and Eq.~\eqref{rot_sign_2}. It is assumed that the initial number of photons $N_0$ is fixed (i.e., $l$ independent). Maximizing the signal-to-noise ratio with respect to $l$, we obtain the standard condition $l=2L(u,q)$. However, for the cavity experiments, the optimization procedure should be slightly modified. First, note that the intensity of the light coupled inside the cavity (the intracavity intensity) is very high: $I_\text{int}\approx I_0/\delta$~\cite{Mes07}, where $\delta$ is the mirror transmission. Second, the mirror transmission is roughly related to the possible number $N_p$ of light passes along the cavity via $\delta \approx N_p^{-1}=l_0/l$, where $l_0$ is the optical pathlength after one pass of the light through the cavity. Now we assume that $N_0\sim \delta \approx N_p^{-1}=l_0/l$, so the corresponding optimization condition follows from the equation
\begin{equation}
\frac{d}{dl} \left [ l \left(\frac{l_0}{l}\right)^{1/2} \text{e}^{-l/2L(u,q)} \right ]=0.
\end{equation}
From this equation, it follows that the optimization for the optical cavities yields $l = L(u,q)$, which differs from the standard optical optimization condition by a factor of $2$.

Now we are able to compare the resonance and off-resonance scenarios. Comparing Eqs.~\Br{rot_sign_0} and~\Br{rot_sign_large}, one reveals the advantage of the optimized experiment at large detunings over the resonance case:
\begin{equation}
\left.\frac{R(u\gtrsim 4)}{R(u=0)} \right|_{\text{opt}} = \frac{\Gamma_{\text{D}}}{\Gamma} \gg 1.
\end{equation}
For the proposed experiment with ThO and PbF molecular beams, $\Gamma_{\text{D}}/\Gamma\sim (10^4 - 10^5)$ (see below). However, the absorption length at large detunings may exceed the resonance-setup value by several orders of magnitude [$L(u\gtrsim 4,q\ll 1)/L(0,q\ll 1)=u^2/q$]. To achieve such large values of $L$ in the off-resonance experiments, one has to take advantage of optical cavities, which allows one to significantly increase the optical pathlength. 

It turns out that further optimization (that beyond the condition $l = L(u,q)$, which is now implied) can be applied if we fix the number of molecules $N$ and take into account the saturation effects in order not to ``bleach'' the molecules. Although we wish to have as many photons ($N_{\text{ev}}$) as possible, this kind of optimization yields the condition $N_{\text{ev}}\approx N$ (see the discussion in Ref.~\cite{Bud20}). This brings us to the so-called ``Equation One'' in the polarimetry~\cite{Bud20}, which in terms of the uncertainty in the electron EDM value $\delta d_e$ for an ideal molecular-beam experiment takes the form
\begin{equation}
\label{Eq1}
\delta d_e \sim \frac{1}{\mathcal{E}_{\text{eff}}} \times  \frac{1}{\tau_c} \times \frac{1}{\sqrt{\dot{N}_{\text{ev}} T}}.
\end{equation}
This is the figure-of-merit for the fundamental noise-limited experiment which determines the statistical sensitivity of the experiment.

Let us now compare the sensitivity of the electron EDM experiments by means of Eq.~\Br{Eq1}. For the electron spin-precession experiment (e.g., such  as that with ThO~\cite{ACME18}), the coherence time $\tau_c$ amounts to a few microseconds, $\dot{N}_{\text{ev}} $ is the number of {\it molecules} per unit time in the desired state and $T$ is the total time of the experiment (usually, about two weeks). In our scheme (optical rotation experiment), from Eqs.~\Br{rot_sign_large} and \Br{R-to-F} for the absorption length, it follows that $\tau_c = 1/\Gamma$ also yields a few microseconds. The quantity $T$ is still the total time of the experiment (usually, about two weeks). Nevertheless, $\dot{N}_{\text{ev}}$ is the number of detected {\it photons} per unit time. This fact is the key difference between our proposal and the electron spin-precession experiments.

On the one hand, as was pointed out above, the optimized setup implies that the number of photons is approximately the same as the number of molecules, $N_{\text{ev}}\approx N$. However, it should be noted that in the existing cavities~\cite{Boug14,Baev99} it is not possible to achieve the absorption length in the case of strongly detuned light for the molecular systems of interest (ThO, PbF), i.e., in fact, $l < L(u,q)$. Therefore, our proposal is not optimal in this sense. In this case, Eq.~\eqref{Eq1} should be modified as follows:
\begin{equation}
\label{Eq1_2}
\delta d_e \sim \frac{L(u,q)}{l} \times \frac{1}{\mathcal{E}_{\text{eff}}} \times  \frac{1}{\tau_c} \times \frac{1}{\sqrt{\dot{N}_{\text{ev}} T}},
\end{equation}
and the number of detected photons can be much larger than the number of molecules, so our goal is to make the number of the former {\it as large as possible in order to reduce the last factor in Eq.~\eqref{Eq1_2}}. Accordingly, the aim of this paper is to evaluate the maximum possible number of detected photons and sensitivity to the electron EDM measurement in the proposed scheme taking into account the transit-time effects, which will be done next.

Finally, let us estimate the maximal number of photons $N_\text{ev}$ in terms of the maximal intracavity intensity $I_\text{max}$. We define the latter quantity by requiring that the absorption probability amount to one tenth,
\begin{equation}
    \label{max_int}
  \left.  \overline{W}_{\Delta}(\tau, \omega_\text{L})\right|_{I=I_\text{max}}
      =0.1.
\end{equation}
A promising candidate for the $\mathcal{P}$,~$\mathcal{T}$-odd Faraday experiment in an optical cavity with diatomic molecules is a PbF molecule with the X1$^2\Pi_{1/2} \rightarrow$ X2$^2\Pi_{3/2}$ transition ($\lambda=1210$~nm). The natural linewidth of the X2 state is $\Gamma=2.7 \times 10^3$~s$^{-1}$~\cite{Das02}. For the PbF beam, we adopt the transverse temperature of 1~K (e.g., in Ref.~\cite{Alm17} the transverse temperature of the supersonic YbF beam was reported to be about 1~K) and $\Gamma_{\rm D}= 4.5 \times 10^{7}$~s$^{-1}$. Let us estimate the maximum intracavity intensity for the transition under investigation in the PbF molecule. According to \Eq{max_int}, for $\tau=10^{-5}~\text{s}$, $\omega_{\rm L}=1.56\times 10^{15}$~s$^{-1}$, and $u=5$, one obtains
\begin{equation}
    \label{max_int_PbF}
   I_{\rm max} \approx 4 \times 10^2~{\rm W}/{\rm cm}^2.
\end{equation}
According to \Eq{sat_int}, the saturation intensity for the above-mentioned parameters in PbF is $I_{\text{sat}}=5.3 \times 10^3~\text{W/cm}^2$.

Consider now the X$^1\Sigma_0\rightarrow$ H$^3\Delta_1$ transition ($\lambda=1810$~nm) in ThO. The natural linewidth of the metastable H state is $\Gamma=2.38\times 10^{2}$~s$^{-1}$~\cite{Ang22}. Substituting the parameters of ThO ($\tau=10^{-5}~\text{s}$, $\omega_{\rm L}=1.04\times 10^{15}$~s$^{-1}$, $\Gamma_{\rm D}=2.9\times10^7$~s$^{-1}$, and $u=5$) in \Eq{max_int}, one finds  
\begin{equation}
    \label{max_int_ThO}
   I_{\rm max} \approx 6.6 \times 10^3~{\rm W}/{\rm cm}^2.
\end{equation}
According to \Eq{sat_int}, the saturation intensity in ThO is $I_{\text{sat}}=7.4 \times 10^3~\text{W}/\text{cm}^2$.
One observes that the condition $I_{\rm max} \gg I_{\text{sat}} $ is not fulfilled. This obviously stems from the fact that even though $\Gamma\tau \ll 1$, the factor 50 in \Eq{prob_sat} can violate this inequality for actual experimental parameters. Note, however, that in the stationary treatment of the problem one should use the laser intensity $I\ll I_{\text{sat}}$ in order to keep the absorption probability sufficiently small.

Finally, taking into account the maximum possible number of the detected photons [Eqs.~\Br{max_int}--\Br{max_int_ThO}], we are going to estimate the expected sensitivity of the electron EDM measurement [\Eq{Eq1_2}] via the $\mathcal{P}$, $\mathcal{T}$-odd Faraday effect in optical cavities. The effective electric field for the excited state of interest in the PbF molecule $\mathcal{E}_{\text{eff}} (^2\Pi_{+3/2}) = 9.3$~GV/cm~\cite{Chub19:3} and in the ThO molecule $\mathcal{E}_{\text{eff}} (^3\Delta_1) = 80$~Gv/cm~\cite{Skrip16}. Recall that $\tau_c = 1/\Gamma$, $T$ is of the order of two weeks and adopt the diameter of the laser beam of the order of 1~mm$^2$. Substituting these parameters in\Eq{Eq1_2}, one finds that the current electron EDM sensitivity can be improved by 1-2 orders of magnitude: $d_e \sim 10^{-32}$ $e$cm even for not fully optimized experimental scheme proposed here. With respect to the possible new-physics effects, this allows one to test new particles at energy one order-of-magnitude larger than the current best constraint.

Furthermore, one may also expect that the pathlength $l$ will be increased in the near future due to the progress in the optical cavity techniques. In this case, adopting the number density and the cross-section of the molecular beams as $\rho\sim 10^{10}$~cm$^{-3}$ and 1 cm$^2$, respectively, and substituting them into~\Eq{Eq1} [for $l=L(u,q)$, $N_{\text{ev}}$ is approximately equal to the number of molecules that have interacted with the laser light], one can improve the current electron EDM sensitivity by 3-4 orders of magnitude ($d_e \sim 10^{-34}$ $e$cm). This value is very close to the benchmark Standard-Model prediction for the $\mathcal{P}$,~$\mathcal{T}$-odd effects $d_e^{\text{eqv}} \sim 10^{-35} \; e\text{cm}$~\cite{Posp22}. The equivalent electron EDM $d_e^{\text{eqv}}$ is defined as the electron EDM which produces the same linear Stark shift in the same electric field as that produced by the electron-nucleus $\mathcal{P}$,~$\mathcal{T}$-odd interaction in a particular atomic system.

In the optimal case $l=L(u,q)$, the sensitivity $\delta d_e$ is determined by means of Eq.~\eqref{Eq1} both for our proposal and for the spin-precession experiment, so one has to explicitly indicate the origins of the 3-4-order-of-magnitude improvement revealed above. First, within our experimental scheme, one can achieve much larger numbers of molecules (and accordingly, photons) since one does not need to prepare molecules in the desired electronic state and the experiment involves the particles in the ground state. Second, in our case, the carrier of the effect is a photon, so it does not matter what happens with a molecule after the interaction process. In the electron spin-precession experiments, from the instant of the injection of molecules (after interaction with external fields and before readout procedure), one loses a lot of molecules, which makes the statistics of the experiment worse.

\section{Conclusions}

In this paper, we evaluated the photon absorption probability in the dipole approximation within short-time interactions between atoms (molecules) and laser radiation. In particular, we were interested in the setup where atoms traverse a continuous laser beam inside an optical cavity with a small transit time. Our investigation was motivated by the question whether the absorption probability (and, accordingly, the population of the excited state of the two-level atomic system) is reduced in the case of short-time interactions. This is extremely important for our proposal put forward earlier and concerning the possible search for the electron EDM in optical cavities. Here we performed a detailed quantitative analysis of the absorption process and obtained more accurate numerical estimates. It is demonstrated that the reduction can be substantial ($W_0\sim \tau^2$) in the resonant non-Doppler case (if $\Gamma_{\text{D}} \tau \ll 1$). In the case of a large detuning with account for Doppler broadening, which is most relevant to the proposed EDM experiments~\cite{Chub21}, we refined the rough estimates for the absorption probability obtained previously. Although the reduction effect of smaller transit times within the scenario involving Doppler broadening vanishes, the advantage over the stationary scenario is in the following. In the finite-transit-time problem, one can have the intracavity intensity on the level of the saturation intensity, $I\sim I_{\rm sat}$, while in the stationary problem one should make sure that $I\ll I_{\rm sat}$. Using the results obtained, the expected electron EDM sensitivity can be improved by 1-2 orders of magnitude ($d_e \sim 10^{-32}$ $e$cm even for not fully optimized experimental scheme). However, further optimizing the experiment by refining the optical-cavity techniques and increasing the optical pathlength, the current electron EDM sensitivity can be improved by 3-4 orders of magnitude ($d_e \sim 10^{-34}$ $e$cm).

\begin{acknowledgments}
This work was supported by the Russian Science Foundation (Grant No. 22-12-00258). D.V.C. also acknowledges the support of the President of Russian Federation Stipend No. SP1213.2021.2. 
\end{acknowledgments}

\appendix*

\begin{widetext}

\section{Integration over $k_0$ in \Eq{prob_2}}

First, let us introduce $x = (\sqrt{2}/4) (k_0-\omega_{\rm L})\tau$. The integral will then have the following form:
\begin{equation}
    \label{appA_prob}
W= \frac{16\pi^2\alpha^2\tau^2 \omega_{\rm L}^2I }{3}  \big| \boldsymbol{e}_{\rm L}\boldsymbol{r}_{{\rm eg}} \big|^2 \big|\boldsymbol{r}_{{\rm ge}}\big|^2 \int\limits_{-\frac{\omega_{\rm L}\tau}{2\sqrt{2}}}^\infty \frac{ {\rm e}^{-x^2} \big (x+\frac{\omega_{\rm L}\tau}{2\sqrt{2}}\big )}{\big (x+\frac{\Delta\omega\tau}{2\sqrt{2}}\big )^2+ \frac{(\Gamma\tau)^2}{32}} \, dx,
\end{equation}
where $\Delta \omega=\omega_{\rm L}-\omega_0$ is the detuning from the atomic resonance. Since the characteristic time for the atom to traverse the laser beam is $\tau\sim 10^{-5}$~s, for visible or near-infrared light $\omega_{\rm L}\tau \gg 1$. Note also that the presence of the exponential factor (${\rm e}^{-x^2}$) in the integrand allows one to extend the lower limit of the integration to $-\infty$ and neglect $x$ in the numerator. It yields
\begin{equation}
    \label{appA_prob2}
W= \frac{2^{5/2}\pi^2\alpha^2\tau^3 \omega_{\rm L}^3I }{3}  \big| \boldsymbol{e}_{\rm L}\boldsymbol{r}_{{\rm eg}} \big|^2 \big|\boldsymbol{r}_{{\rm ge}}\big|^2 \int\limits_{-\infty}^\infty \frac{{\rm e}^{-x^2}dx}{\big (x+\frac{\Delta\omega\tau}{2\sqrt{2}} \big )^2+ \frac{(\Gamma\tau)^2}{32}}.
\end{equation}
It is convenient to express the result in terms of the real-valued function $f$ defined in \Eq{fg}. This function has the following integral representation~\cite{Khrip91}:
\begin{equation}
\label{faddeeva_int}
\frac{1}{\sqrt{\pi}} \int \limits_{-\infty}^{\infty} dx \, \frac{{\rm e}^{-x^2}}{x - u -  iq} = if(u, q) - g(u, q),
\end{equation}
where $g(u, q)$ is also assumed to be real. The integral in \Eq{appA_prob2} is then trivial:
 \begin{equation}
\label{int1}
 \int \limits_{-\infty}^{\infty} \frac{{\rm e}^{-x^2}dx}{(u-x)^2+q^2} = - \frac{1}{q} \, {\rm Im}\int \limits_{-\infty}^{\infty} \frac{{\rm e}^{-x^2}dx}{u-x+iq}= \frac{\sqrt{\pi}}{q} \, f(u,q).
\end{equation}
Replacing $x$ with $-x$ in \Eq{appA_prob2} leads us to the integral of the form~\Eq{int1}. Finally, we obtain
\begin{equation}
    \label{appA_prob3}
W= \frac{32\pi^{5/2} \alpha^2\tau^2 \omega_{\rm L}^3I }{3\Gamma}  \big| \boldsymbol{\varepsilon}_{\rm L}\boldsymbol{r}_{{\rm eg}} \big|^2 \big|\boldsymbol{r}_{{\rm ge}}\big|^2 f\left(\frac{\Delta\omega\tau}{2\sqrt{2}},\frac{\Gamma\tau}{4\sqrt{2}}\right),
\end{equation}
which coincides with \Eq{prob3}.

\end{widetext}

\end{document}